\documentclass[12pt]{article}

\usepackage{epsfig}

\newcount\xrefpos \xrefpos=0
\newcount\yrefpos \yrefpos=0
\newcount\xput \xput=0
\newcount\yput \yput=0

\def\refpos#1 #2 #3{\global\xrefpos=#1 \global\yrefpos=#2
                         \rlap{$\smash{#3}$}}
\def\put #1 #2 #3{\xput=#1 \yput=#2
                  \advance\xput by -\xrefpos
                  \advance\yput by -\yrefpos
                  \rlap{\kern\the\xput truebp
                        \vbox to 0pt{\vss\hbox{$\displaystyle #3$}
                        \kern\the\yput truebp}}}
\def\beginlabels\refpos#1\endlabels{\hbox{$\refpos#1$}}

\newcommand{\lb}{\langle}
\newcommand{\rb}{\rangle}
\newcommand{\sla}{\!\!\!/}
\newcommand{\nn}{\nonumber}
\newcommand{\eps}{\epsilon}
\newcommand{\veps}{\varepsilon}

\newcommand{\la}{\ensuremath{\lambda}}

\newcommand{\A}{\ensuremath{{\cal A}}}

\newcommand{\half}{\ensuremath{\frac{1}{2}}}

\newcommand{\be}{\begin{equation}}
\newcommand{\ee}{\end{equation}}
\newcommand{\bea}{\begin{eqnarray}}
\newcommand{\eea}{\end{eqnarray}}

\begin{document}

\begin{titlepage}

\bigskip
\hskip 4.8in\vbox{\baselineskip12pt \hbox{hep-th/0410278}}
\newline
\hskip 4.8in\vbox{\baselineskip12pt \hbox{NSF-KITP-04-119}}

\bigskip
\bigskip
\bigskip

\begin{center}

{\Large \bf  One-Loop MHV Amplitudes in Supersymmetric Gauge
Theories}
\end{center}

\bigskip
\bigskip
\bigskip

\centerline{\bf Callum Quigley$^a$  and Moshe Rozali$^{\ a, b}$}

\bigskip
\bigskip
\bigskip

\centerline{$^a$\it Department of Physics and Astronomy} \centerline{\it
University of British Columbia} \centerline{\it Vancouver, British
Columbia V6T 1Z1, Canada} \centerline {} \centerline {$^b$ \it Kavli Institute
for Theoretical Physics}\centerline{\it University of California}
\centerline{\it Santa Barbara, CA 93106-4030} \centerline{}\centerline{\small \tt cquigley,
rozali@physics.ubc.ca}

\bigskip
\bigskip

\begin{abstract}
\vskip 2pt Using CSW rules for constructing scalar Feynman diagrams
from MHV vertices, we compute the contribution of $\mathcal {N}=1$
chiral multiplet to  one-loop MHV gluon amplitude. The result agrees
with the one obtained previously using unitarity-based methods,
thereby demonstrating the validity of the MHV-diagram technique,
in the case of one-loop MHV amplitudes, for all massless supersymmetric theories.

\end{abstract}

\end{titlepage}


\baselineskip=18pt \setcounter{footnote}{0}

\section{Introduction and Summary}

Helicity amplitudes in gauge theories with massless matter exhibit
remarkable simplicity, most manifest in the maximal helicity
violating (MHV) amplitudes. For external gluons those are amplitudes
with two negative helicity gluons and any number of positive
helicity gluons (or vice versa). At tree level they are given by the
Parke-Taylor formula, \cite{pt} which is holomorphic in the spinor
variables characterizing the momenta of the external gluons. Such
simplicity hints at a large symmetry structure which is hidden in
the usual formulation of perturbation theory.

The simplification is not limited to MHV amplitude at tree level. In
\cite{witten} a precise formulation\footnote{See \cite{nair} for
earlier related work.} of the structure of helicity amplitudes was
given: when transformed to twistor space, the helicity
amplitudes are supported on certain algebraic curves. The precise
type of algebraic curve depends on the detail of the amplitude, and
grows more complicated with increasing number of loops or decreased
helicity violation. The simplest such amplitudes are the
Parke-Taylor ones, which are supported on lines in twistor space.

This structure led \cite{csw} to formulate effective Feynman rules
for obtaining arbitrary helicity amplitudes using the  Parke-Taylor
amplitudes (continued to off-shell momenta) as vertices, combined
with simple scalar propagators. We will call such Feynman graphs
MHV-diagrams. Already in \cite{csw}  a strong case was made that
tree level helicity amplitudes are reproduced using these rules, and
this was confirmed in a series of papers \cite{tree}, for a review
see \cite{review}. Nevertheless a direct derivation of these rules
from  field theory is still lacking (see however \cite{nairnew}).

The simple structure in twistor space motivates exploration of
string theories with twistor target space \cite{witten, etc}. The
original string theory in \cite{witten} indeed reproduces the tree
level amplitudes \cite{strings} (the relation of these calculations
to  MHV-diagrams was explained in \cite{explanation}). However the
correspondence with the $\mathcal{N}=4$ amplitudes breaks down at
one loop \cite{berko}, due to non-decoupling of conformal
supergravity modes.

The twistor space structure of one-loop diagrams  was initially less
clear, as they seem to be supported in twistor space on
configurations different from  one-loop  MHV-diagrams \cite{csw2}.
Nevertheless, application of the MHV-diagram formalism in \cite{bst}
reproduces the known result for one-loop MHV amplitude in $\mathcal
{N}=4$ theory. The discrepancy was clarified by the existence of an
holomorphic anomaly \cite{holomorphic}, which can be furthermore
used to calculate unitarity cuts \cite{moreholo}. For further
discussion of one-loop amplitudes in the $\mathcal{N}$=4 theory see
\cite{more}.

So far all the one-loop results are restricted to the maximally
supersymmetric case. As the formalism of MHV-diagrams lacks a field
theory derivation, it is necessary to check it by reproducing the
known one loop results, before using this efficient formalism to
calculate unknown amplitudes.  In this note we compute the
contribution of the $\mathcal{N}=1$ chiral multiplet to the MHV one
loop amplitudes. All MHV one-loop amplitudes in massless supersymmetric
theory are a linear combination of this contribution, and that of a
vector multiplet of $\mathcal{N}=4$ SYM, calculated in \cite{bst}. We
therefore confirm that the MHV-diagram technique works for any
supersymmetric theory at one-loop, at least for the amplitude we discuss. It seems that the success hinges
more on the cut-constructibility  \cite{fusing}, rather than
supersymmetric cancelations. It would be interesting to check
further whether this formalism is valid  for non-supersymmetric (but
still cut-constructible) cases.

This note is organized as follows: in section 2 we describe the CSW
rules for constructing MHV diagrams \cite{csw}, and their
application to one-loop calculations in \cite{bst}. We also present
the $\mathcal{N}=1$ amplitude constructed in \cite{fusing},
introducing our notations in the process. Section 3 includes our
calculation, we follow closely the methods of \cite{bst}: we start by evaluating
the MHV one-loop diagrams, we then arrange the result according to its cuts, and finally we
perform
the dispersion integration, which reproduces the amplitude from its cuts.

\section{Background}

\subsection{One-Loop MHV Diagrams}

The CSW rules \cite{csw} for constructing MHV diagrams consist of
using the MHV amplitudes with two negative helicity gluons, as well
as their supersymmetric partners, as the basic building blocks for
obtaining all amplitudes (including the so-called googly ones, which
have two positive helicity gluons, and therefore are also MHV).

Denote all incoming momenta into a vertex by $k_i$. The CSW
prescription associates with each such momentum a spinor $\la_i$.
For null momenta, $k_i^2=0$ the assignment follows from the
decomposition $k_i^\mu = (\sigma^\mu)_{\alpha \dot{\alpha}}
(\la_i)^\alpha (\bar{\la_i})^{\dot{\alpha}}$. For general massive
momenta this decomposition is impossible. Instead one chooses
arbitrary spinors $\eta, \bar{\eta}$, equivalent to the choice of a
lightcone frame, and decomposes each off-shell momentum $k_i$ as \be
(k_i)^{\alpha \dot{\alpha}} = (\la_i)^\alpha
(\bar{\la_i})^{\dot{\alpha}} + z_i (\eta)^\alpha
(\bar{\eta})^{\dot{\alpha}} \ee The new variables $z_i$ express how
virtual   are the momenta $k_i$.

Having chosen spinors $\la_i$ for all momenta $k_i$, the vertices in
the diagrams are the holomorphic Parke-Taylor amplitudes. When
labeling all momenta cyclically, in clockwise direction, the purely
gluonic vertices\footnote{We introduce the supersymmetric partners
of these vertices in the next section.} are \be \lb p\ q \rb^4
\prod_{i=1}^n \frac{1}{\lb i, i+1\rb} \ee where $n$ is the number of
gluons in the vertex (which can be arbitrary), the negative helicity
gluons are in positions $p,q$, and we use the notation $ \lb i\ j
\rb \equiv \epsilon_{\alpha \beta} (\la_i)^\alpha(\la_j)^\beta$ for
any pair of spinors. For later use we also define $  [i \,j]= \equiv
\epsilon_{\dot{\alpha} \dot{\beta}}
(\bar{\la_{i}})^{\dot{\alpha}}(\bar{\la_{j}})^{\dot{\beta}}$ and
$\lb \la \,|P\,|\,\bar{\eta}\,]= P_{{\alpha}
\dot{\beta}}(\la)^\alpha (\bar{\eta})^{\dot{\beta}}$.

The MHV vertices are connected by simple scalar propagators,
$\frac{i}{L^2}$, for any off-shell momentum $L$. As explained  in
\cite{bst}, when combined with the integration over all off-shell
momenta, the result is independent of the lightcone direction
chosen, and can be decomposed as \be \frac{d^4 L}{L^2} = 2i
\,{d^4\ell \,\delta^{(+)}(\ell^2)}\,\frac{dz}{z} \label{dlips}\ee
where $\ell= \la \bar{\lambda}$ is the null momentum associated to
$L$ by the CSW prescription, and $\delta^{(+)}(\ell^2)$ is precisely
the Lorenz invariant phase space measure appearing in the
calculation of unitarity cuts. The decomposition is crucial to the
results of \cite{bst} and will play the same role for us: the
appearance of the Lorenz invariant phase space measure allows us to
use unitarity-based methods \cite{fusing}. The final integration
over the z-variables is a dispersion integration which reconstructs
the amplitude from its cuts.

\subsection{The $\mathcal{N}=1$ Amplitude}

The contribution of the  $\mathcal{N}=1$ chiral multiplet to
one-loop MHV amplitude was calculated in \cite{fusing} using
unitarity-cut  methods. We quote here the result we reproduce later
using the MHV one-loop diagrams. Our notation here is fairly similar
to \cite{csw2}, who analyzed the twistor space structure of this
amplitude.

The result obtained in \cite{fusing} is then: \bea
\label{result}\A_{chiral} = \frac{\A_{tree}}{32\pi^2}\, \left\{
\sum_{r,s} b_{rs}^{pq}\, B(k_r,Q, k_s, P) +\sum_{r,s} c_{rs}^{pq}\,
T(k_r,P,\widetilde{Q})+ \sum_{r,s} c_{sr}^{pq}\,
T(k_s,Q,\widetilde{P}) +\A_{IR}\right\} \eea where $\A_{tree}$ is
the tree level MHV amplitude, and $p,q$ are the locations of the
negative helicity gluons. The first term in the parentheses comes
from scalar (2-mass) box diagram, then there are two terms coming
from scalar (two-mass) triangles and finally the last part $\A_{IR}$
comes from exceptional, boundary terms.  The explicit form of the
functions and some of their properties are summarized in  appendix
II of \cite{fusing}, and is also summarized below.

 We now explain our notation in
formula (\ref{result}), including  the ranges of all summations.
First, the box functions $B(k_r,Q,k_s,P)$ are the finite part of
those appearing in the $\mathcal{N}=4$ theory.  Using the
representation discovered in \cite{bst}, they are \bea\label{box}
B(k_r,Q,k_s,P) &=&
F(k_r,Q,k_s,P)+\frac{1}{\eps^2}\mbox{\LARGE[}(-s)^{-\eps}+
(-t)^{-\eps}-(-P^2)^{-\eps}-(-Q^2)^{-\eps}\mbox{\LARGE]}\nn\\
&=&
\mbox{\rm{Li}}_2(1-aP^2)+\mbox{\rm{Li}}_2(1-aQ^2)-\mbox{\rm{Li}}_2(1-as)-\mbox{\rm{Li}}_2(1-at)^2,\eea
where we have introduced the momentum invariants $s=(P+k_{r})^2$ and
$t=(P+k_{s})^2$ and the quantity $a$ is defined \be
a=\frac{P^2+Q^2-s-t}{P^2Q^2-st}.\ee These functions are
characterized by two massless external legs, with momenta $k_r,k_s$.
The remaining momenta are then arranged uniquely into two massive
external legs with momenta $P,Q$. The range of summation over $r,s$
in (\ref{result}) is restricted such that $p$ belongs to the set of
momenta in $P=k_{s+1}+\ldots+k_{r-1}$, and likewise $q$ is one of
the momenta in $Q=k_{r+1}+\ldots+k_{s-1}$. In particular the
massless momenta $k_r,k_s$ always have positive helicity. The set of
diagrams contributing to the box functions is drawn in figure 1.

\begin{figure}[ht]
\begin{center}
\epsfbox{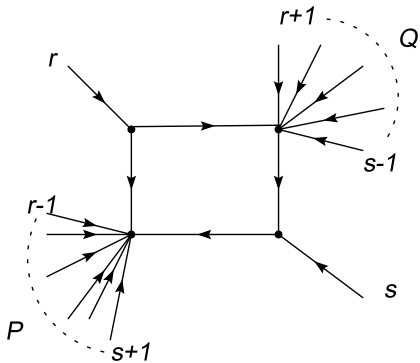}
         \end{center}
         \caption{Diagrams contributing to box functions.}
\end{figure}
The triangle diagrams have similarly one massless momentum
$k_{r,s}$, and two massive ones: $P,\widetilde{Q}\equiv Q+k_s$ or
$Q,\widetilde{P}\equiv P+k_r$, each of
 which contains a single negative helicity gluon.
 Furthermore, to posses two massive legs requires $|r-s|>1$
and $|r-s-1|>1$. The form of the two triangle functions are
identical, in general \be T(k,P,Q)=
\frac{\log(P^2)-\log(Q^2)}{P^2-Q^2}\ee The two ranges of summations
only differ from those of the box functions in that the first sum
includes $s=q$ and the second $r=p$. Note also that the coefficients
$c_{rs}^{pq}$ depend on which leg is null, as explained below. The
set of diagrams contributing to triangle functions
$T(k_r,P,\widetilde{Q})$ is drawn in figure 2, the others follow the
same pattern.
\begin{figure}[ht]
\begin{center}
\epsfbox{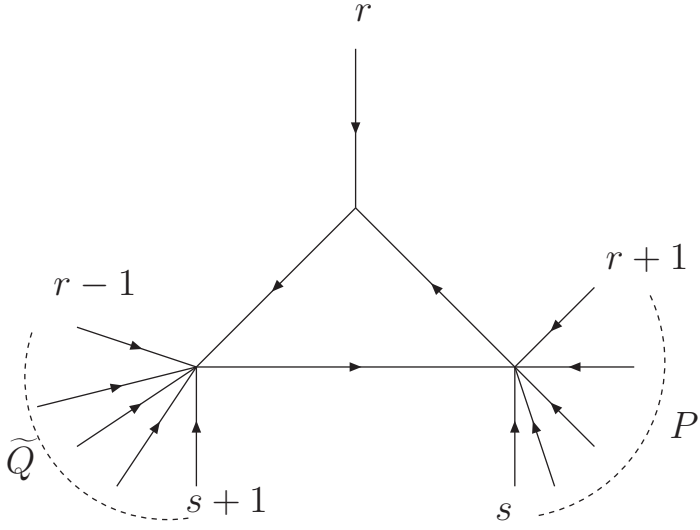}
         \end{center}
         \caption{Diagrams contributing to triangle functions.}
\end{figure}

The last term $\A_{IR}$ arises from degenerations of triangle
diagrams for which one of the massive momenta become massless, that
is it contains only a single external momentum which is then
necessarily a negative helicity gluon. There are four such
degenerations, for which $p=P,\widetilde{P}$ or $q=Q,\widetilde{Q}$.
These cases are drawn in figure 3, they give rise to the following 4
terms: \be \A_{IR} = c_{p+1,p-1}^{pq}
\frac{\left(-(k_{p+1}+k_{p-1})^2\right)^{-1-\eps}}{\eps(1-2\eps)} +
c_{p-1,p}^{pq}\frac{\left(-(k_{p-1}+k_{p})^2\right)^{-1-\eps}}{\eps(1-2\eps)}
+ \left(p\leftrightarrow q \right). \ee

\begin{figure}[h]
\begin{center}
\epsfbox{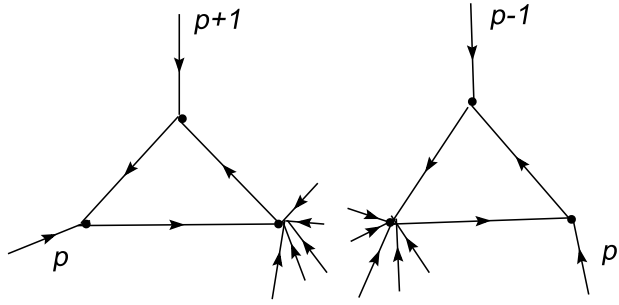}
         \end{center}
         \caption{Two of the degenerate triangle diagrams, the other two are obtained
                       by   exchanging  $p$ and $q$ .}
\end{figure}
Finally, the coefficients appearing in (\ref{result}) are as
follows. For the box functions \be
 b_{rs}^{pq} = 2 \frac{\lb p \ r \rb \lb p \ s\rb \lb q \ r\rb
 \lb q \ s \rb }{\lb r \ s \rb^2 \lb p \, q \rb ^2}\label{bpqrs}
\ee whereas for the triangles (and the boundary terms) one has \be
c_{rs}^{pq}= \frac{\lb p \ r \rb \lb r \ q \rb}{\lb p \ q \rb^2}
\frac{\lb s,s+1\rb }{\lb s \ r \rb \lb s,r+1 \rb} \mbox{\LARGE(}\lb
q\ r\rb \lb p\,|P\,|r]+ \lb p \ r\rb \lb q\,|P\,|r]\mbox{\LARGE)}.
\ee  Notice for $c_{sr}^{pq}$ we must change
$P=k_{s+1}+\ldots+k_{r-1}$ to $Q=k_{r+1}+\ldots+k_{s-1}$.

\section{The MHV Diagram}

\subsection{Reduction of the Diagram}

 We are interested in calculating the contribution of $N=1$ chiral
 multiplet to one-loop MHV amplitudes.
 We consider the case of external
 gluons only, but many other diagrams with external fermions or scalar
 are related to this amplitude by supersymmetry. The one-loop diagram we compute is MHV, having two
 negative helicity gluons and arbitrary number of positive helicity
 ones.

 The typical one loop MHV diagram of interest is drawn in figure 4.
 We must have one negative helicity gluon on each side of the diagram, as
 there is no possible helicity assignment for the intermediate states if both
 negative helicity gluons are on the same side of the diagram. We
 label the momenta on the left side as $k_{m_1},...,k_{m_2}$, one of which is negative helicity,
 denoted by $p$. The momenta  on the right side as $k_{m_2+1},..., k_{m_1-1}$, the negative
 helicity momentum labeled $q$. All momentum
 labels are cyclically ordered. When calculating the complete
 amplitude one has to sum over such MHV diagram, we will arrange
 this sum according to the cuts, following \cite{bst}. All loop momenta
 are evaluated using dimensional regularization, in $D$
 dimensions, with $D=4-2\eps$.

 \begin{figure}[ht]
\begin{center}
\epsfbox{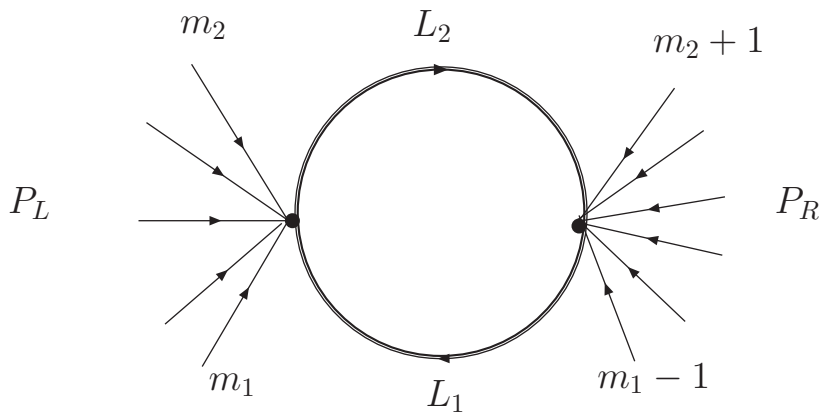}
         \end{center}
         \caption{Typical one-loop MHV diagram, one has to sum over all choices of $m_1,m_2$.
         The negative helicity gluon $p$ is on the left, and $q$ is on the right.}
\end{figure}

 Now the amplitude for the diagram in figure 4  is given
by\be A_{chiral}^{1-loop} = i (2\pi)^4 \delta(P_L+P_R) \int
\frac{d^4 L_1}{L_1^2}
 \int \frac{d^4 L_2}{L_2^2} \,\delta^{(4)}(L_1-L_2+P_L)\left( A_L^{F} A_R^{\tilde{F}} +
 A_L^{\tilde{F}} A_R^{F}+ 2
 A_L^{S} A_R^{S} \right)\ee
where $P_L, P_R$ are the momenta
flowing into the diagrams from the
 left  and right correspondingly. The factors $A_L, A_R$ are related by supersymmetry
 to the gluonic MHV introduced above. Each
 vertex is obtained by combining the external gluons with two
 internal lines, which are members of a chiral multiplet, including a
 fermion (of two helicities, resulting in vertices $A^{F}$ and $A^{\tilde{F}}$)
 and a complex scalar (resulting in a vertex $A^{S}$).

 As reviewed above, each of the off-shell momenta $L_i, i=1,2$ can be
 associated a null momentum $\ell_i$ and the corresponding spinors $\ell_i= \la_i \bar{\la_i}$.
 These are necessary in order to write the off-shell vertices. We
 have therefore $L_i= \ell_i +z_i \eta \bar{\eta}$.

  Factoring out the tree (Parke-Taylor) amplitude results\footnote{We do
  not keep track of the overall
  sign, which can be fixed at the end of the calculation.} in:
\bea \label{for1} A_{chiral}^{1-loop} &=& A_{tree} \,\int \frac{d^4
L_1}{L_1^2} \int \frac{d^4 L_2}{L_2^2}
 \,\delta^{(4)}(L_1-L_2+P_L) \frac{1}{\lb \la_1 \ \la_2\rb \lb \la_2 \ \la_1\rb} \left( 2I^{S} + I^{F}+ I^{\tilde{F}}\right)
   \nonumber\\
  &\times& \frac{\lb m_2 \ m_2+1\rb \lb m_1-1 \ m_1\rb}
  {\lb \la_1 \ m_1\rb  \lb m_2 \ \la_2\rb \lb \la_2  , m_2+1 \rb \lb m_1-1 , \la_1\rb}
 \eea
where \bea I^{S} &=& \frac{\lb \la_1 \ p\rb ^2 \lb \la_2 \ p \rb ^2
\lb \la_1 \ q\rb^2
\lb \la_2 \ q\rb ^2}{\lb p \ q\rb ^4} \nonumber \\
            I^{F} &=& - I^{S} \frac{\lb \la_2 \ q \rb \lb \la_1 \ p\rb}
            {\lb \la_2 \ p \rb \lb \la_1 \ q \rb} \nonumber \\
            I^{\tilde{F}} &=& - I^{S}
            \frac{\lb \la_2 \ p\rb \lb \la_1 \ q\rb }{\lb \la_2 \ p \rb \lb \la_1 \ q\rb}
            \eea

To sum the 3 terms in (\ref{for1}) one uses the Schouten identity
\be \lb a \ b\rb \lb c \ d \rb =\lb a \ d \rb\lb c \ b\rb +\lb a \ c \rb \lb b \ d\rb \ee which will be repeatedly used
below. This leads to the following expression for the chiral
multiplet contribution to the one loop gluon MHV amplitude \be
A_{chiral}^{1-loop} = A_{tree}\int \frac{d^4 L_1}{L_1^2} \int
\frac{d^4 L_2}{L_2^2}
 \delta^{(4)}(L_1-L_2+P_L)\,\,R
\ee with \be R= \frac{\lb m_1-1,m_1\rb \lb m_2,m_2+1\rb \lb \la_1 \
q\rb \lb \la_2 \ q\rb \lb \la_1 \  p\rb \lb \la_2 \ p\rb}{\lb p \
q\rb ^2 \lb m_1-1,\la_1 \rb \lb \la_1 \ m_1\rb \lb m_2 \ \la_2\rb
\lb \la_2, m_2+1\rb } \ee

Our next step is to split the spinor expression $R$ into 4 terms of
identical structure. Using the pair of Schouten identities: \bea
&\lb m_1-1,m_1\rb \lb\la_1 \ q\rb = \lb m_1-1,q\rb \lb\la_1 \ m_1\rb
+\lb m_1-1,\la_1\rb \lb m_1 \ q\rb&
\nonumber \\
&\lb m_2, m_2+1\rb \lb \la_2 \  p \rb  = \lb m_2 \ p\rb \lb
\la_2,m_2+1\rb + \lb m_2 \ \la_2\rb \lb m_2+1,p\rb& \eea

one gets the following sum: \be R =R(m_2,m_1-1)-
R(m_2+1,m_1-1)-R(m_2,m_1) + R(m_2+1,m_1) \ee where \be R(r,s) =
\frac{\lb \la_1 \ p\rb \lb \la_2 \ q\rb}{\lb p \ q \rb^2} \frac{\lb
s \ q \rb \lb r \ p\rb}{\lb s \ \la_1\rb \lb r \ \la_2\rb } \ee

Let us simplify $R(r,s)$: once again one uses Schouten identities to
split $R(r,s)$ to 4 term, which (when integrated) give rise to
tensor box, triangle and bubble diagrams. The 4 terms  are: \bea
R^A(r,s) &=& \frac{ \lb s \ q\rb \lb r \ p\rb
\lb r \ q\rb \lb p s\rb }{\lb p \ q\rb^2 \lb r \ s\rb^2}
\frac{\lb \la_2 \ s\rb \lb  \la_1 \ r\rb}{\lb \la_1 \ s\rb \lb \la_2 \ r\rb} \nonumber \\
R^B(r,s) &=& \frac{\lb s \ q\rb \lb r \ p\rb^2 \lb r \ q\rb}{\lb p \
q\rb^2 \lb r \ s\rb^2} \frac{\lb \la_2 \ s\rb}{\lb\la_2 \ r\rb}
\nonumber \\
R^C(r,s) &=& \frac{\lb s \ q\rb^2\lb r \ p\rb \lb s \ p\rb}{\lb p \
q\rb^2\lb r \ s\rb^2} \frac{\lb\la_1 \ r\rb}{\lb\la_1 \ s\rb}
\nonumber \\
R^D(r,s)&=& \frac{\lb s \ q\rb^2\lb r \ p\rb^2}{\lb p \ q\rb ^2\lb r \ s\rb^2} \eea

Let us simplify these expressions  one at a time:

\begin{itemize}
\item{\bf Simplifying $R^A$}

We decompose the tensor box function $R^{A}$, as in
\cite{fusing,bst}, into scalar components by expanding
\begin{eqnarray}
\frac{\lb s\ \la_2\rb\lb r\ \la_1\rb}{\lb r\ \la_2\rb\lb s\
\la_1\rb}=\frac{[\la_2\ r]\lb r\ \la_1\rb[\la_1\ s]\lb s\
\la_2\rb}{\lb r\ \la_2\rb[\la_2\ r]\lb s\ \la_1\rb[\la_1\
s]}=-\frac{\textrm{tr}(\frac{1}{2}(1-\gamma^5)\ell\sla_{2}\;
k\sla_r\;\ell\sla_1\;
k\sla_s)}{(\ell_2-k_r)^2(\ell_1+k_s)^2}=\\
=\frac{-2\left\{(\ell_2\cdot k_r)(\ell_1\cdot k_s)+(\ell_2\cdot
k_s)(\ell_1\cdot k_r)-(k_r\cdot
k_s)(\ell_1\cdot\ell_2)+i\veps_{\mu\nu\rho\sigma}
\ell_2^{\mu}k_r^{\nu}\ell_1^{\rho}k_s^{\sigma}\right\}}{(\ell_2-k_r)^2(\ell_1+k_s)^2}\nn
\end{eqnarray}
The term proportional to the $\veps$-tensor vanishes upon
integration. One can define $P_{L;z}= \ell_1-\ell_2= P_L -(z_1-z_2)
\eta \bar{\eta}$ , then the rest of the numerator may be rewritten
as
\bea
\left(2(P_{L;z}\cdot k_r)(P_{L;z}\cdot k_s)-(k_r\cdot
k_s)P_{L;z}^2\right)&-&(\ell_1+k_s)^2(P_{L;z}\cdot
k_r)-\nn \\(\ell_2-k_r)^2(P_{L;z}\cdot k_s) &+&(\ell_2-k_r)^2(\ell_1+k_s)^2
\eea
The terms collected in the first brackets contribute to
a scalar box integral, while the next two terms each contain a
factor which cancels one of the propagators in the denominator
, leaving scalar triangles. The last term reduces to a scalar
bubble, since both propagators cancel.  Next, we make use of the identity
\begin{equation}
4(P\cdot i)(P\cdot j)-2P^2(i\cdot j)=(P+i)^2(P+j)^2-P^2(P+i+j)^2,
\end{equation}
valid for any momentum $P$ and null momenta $i$ and $j$, to rewrite
the box's coefficient in terms the shifted momentum invariants,
defined as $s_z= (P_{L;z})^2,P_z^2= (P_{L;z}-k_r)^2, t_z
=(P_{L;z}-k_r+k_{s})^2$, and $Q_z^2=(P_{L;z}+k_s)^2$:
\begin{equation}
2(P_{L;z}\cdot k_r)(P_{L;z}\cdot k_s)-(k_r\cdot
k_s)P_{L;z}^2=\frac{1}{2}(P_z^2 Q_z^2-s_z t_z)
\end{equation}
Thus, the result of the tensor box's decomposition is
\begin{eqnarray}
\frac{\lb s\ \la_2\rb\lb r\ \la_1\rb}{\lb r\ \la_2\rb\lb s\
\la_1\rb}=\left\{\frac{\half(P_z^2Q_z^2-s_z
t_z)}{(\ell_2-k_r)^2(\ell_1+k_s)^2}-\frac{P_{L;z}\cdot
k_r}{(\ell_2-k_r)^2}-\frac{P_{L;z}\cdot
k_s}{(\ell_1+k_s)^2}\right\}+1\label{tensorbox}
\end{eqnarray}

The  terms collected  in the bracket are the  integrand of a
(divergence free) scalar box function, complete with the correct
coefficient $b_{rs}^{pq}$, as in equation (\ref{result}). The second
term contributes to scalar bubbles, which cancel against other
contributions. We demonstrate this cancelation below.

\item{\bf Simplifying $R^B$ and $R^C$}

We now turn to the linear triangle terms $R^{B}(r,s)$ and
$R^{C}(r,s)$. First, we write the loop momentum dependant part of
$R^{B}(r,s)$ as \be \frac{\lb s\ \la_2\rb}{\lb\la_2\
r\rb}=-\frac{\lb s\ \la_2\rb[\la_2\ r]}{\lb r\ \la_2\rb[\la_2\
r]}=\frac{\lb s|\ell_2|r]}{(\ell_2-k_r)^2}=\lb
s|\gamma_{\mu}|r]\frac{\ell_2^{\mu}}{(\ell_2-k_r)^2}.\ee So, $R^{B}$
is  the integrand of the (cut) linear two-mass triangle integral
$I^{2m}_{3:r-m_1;m_1}[\ell_2^{\mu}]$, defined in \cite{fusing}.
Next, we use the decomposition of the linear triangle given in
\cite{fusing}: \be
I^{2m}_{3:r;i}[k^{\mu}]=-(P_z+k_r)^{\mu}I^{2m}_{3:r;i}[y]-k_r^{\mu}I^{2m}_{3:r;i}[z],\ee
where the arguments in square brackets are the numerators in the
integrals, $y$ and $z$ are Feynmann parameters, $P_z$ is the momentum
of one massive leg (shifted by $z$ dependent terms, as defined above) and $k_r$ is the momentum of the massless leg, as
drawn in figure 3. Since $[r\ r]=0$, we can write \be\lb
s|\gamma_{\mu}|r]I^{2m}_{3:r-m_1;m_1}[\ell_2^{\mu}]=-\lb
s|P_z|r]I^{2m}_{3:r-m_1;m_1}[y].\ee Now, the full coefficient of
$R^{B}$ is\be -\frac{\lb p\ r\rb\lb r\ q\rb\lb p\ r\rb\lb s\ q\rb\lb
s\ P_z\rb[P_z\ r]}{\lb p\ q\rb^2 \lb r\ s\rb^2}.\ee Applying the
Schouten identity to the terms $\lb p\ r\rb\lb s\ P_z\rb$ and $\lb r\
q\rb\lb s\ P_z\rb$, then averaging over the two gives \bea
\label{tri}&-&\frac{\lb p\ r\rb\lb s\ q\rb}{\lb p\ q\rb^2\lb r\
s\rb}\left(\frac{\lb p\ r\rb\lb q\ P_z\rb+\lb q\ r\rb\lb p\
P_z\rb}{2}\right)[P_z\ r]\nn\\ &-&\frac{\lb p\ r\rb\lb s\ q\rb}{\lb p\
q\rb^2\lb r\ s\rb^2}\left(\frac{\lb p\ s\rb\lb r\ q\rb+\lb p\
r\rb\lb s\ q\rb}{2}\right)2(P_z\cdot k_r)\label{tricoeff1}\eea

We use the Schouten identity again, on the first term of the first
pair only. \be \frac{\lb p\ r\rb\lb s\ q\rb}{\lb r\ s\rb}=\frac{\lb
p\ s\rb\lb r\ q\rb}{\lb r\ s\rb}+\lb p\ q\rb\ee   Note that the
piece containing $\lb p\ q\rb$ is independent of $s$, so it will
vanish when summing over $s=\{m_1-1,m_1\}$, as that sum has
alternating signs.  Now the first pair of terms in equation
(\ref{tricoeff1}) reads \be \frac{\lb p\ r\rb\lb r\ q\rb}{\lb r\
s\rb \lb p\ q\rb^2}\left(\frac{\lb s\ q\rb\lb p\ P_z\rb+\lb s\
p\rb\lb q\ P_z\rb}{2}\right)[P_z\ r].\ee A similar analysis of
$R^{C}(r,s)$ shows that the coefficient of the integral function
\newline $I_{3:s-m_2+1;m_2+1}^{2m}[y]$~is \bea & &\frac{\lb p\
s\rb\lb s\ q\rb}{\lb r\ s\rb\lb p\ q\rb^2}\left(\frac{\lb r\ p\rb\lb
q\ Q_z\rb+\lb r\ q\rb\lb p\ Q_z\rb}{2}\right)[Q_z\ s]\nn\\
&+&\frac{\lb p\ r\rb\lb s\ q\rb}{\lb r\ s\rb^2\lb p\
q\rb^2}\left(\frac{\lb p\ s\rb\lb r\ q\rb+\lb p\ r\rb\lb s\
q\rb}{2}\right)2(Q_z\cdot k_s)\label{tricoeff2}\eea where $Q_z$ is
the shifted momentum transfer defined above.
 In this decomposition the first term
contributes to the coefficient of the scalar triangle function, and
the second one will be used below to cancel the bubble diagrams.

\item{\bf Simplifying $R^D$}

First,the scalar bubble in $R^D$ can be combined with that discussed above, in
 $R^{A}$, giving a single bubble with coefficient
\be \frac{\lb p\ r\rb\lb s\ q\rb}{\lb r\ s\rb \lb p\ q\rb^2}\mbox{\LARGE(}\lb p\
s\rb\lb r\ q\rb+\lb p\ r\rb\lb s\ q\rb\mbox{\LARGE)}\nonumber\ee

Now, to cancel the bubbles notice that they posses the same
coefficient as the last pair of terms in equations.
(\ref{tricoeff1}) and (\ref{tricoeff2}).  These integrals combine
into \bea \frac{\lb p\ r\rb\lb s\ q\rb}{\lb r\ s\rb^2\lb p\
q\rb^2}\left(\frac{\lb p\ s\rb\lb r\ q\rb+\lb p\ r\rb\lb s\
q\rb}{2}\right)&\circ&\nn\\
\mbox{\LARGE(}2I_{2:r-m_1;m_1}-2(P_z\cdot
k_r)I^{2m}_{3:r-m_1;m_1}[y]&+&2(Q_z\cdot
k_s)I^{2m}_{3:s-m_2+1;m_2+1}[y]\mbox{\LARGE)},\eea which vanishes in
each channel of each cut because of the relation \be
(t_{i}^{[r+1]}-t_{i}^{[r]})I^{2m}_{3:r;i}[y]=I_{2:r;i}-I_{2:r+1;i}.\ee
Here we have introduced the additional notation
$t_{i}^{[r]}=(k_{i}+k_{i+1}+\ldots+k_{i+r-1})^2$.

\end{itemize}

In summary, the net result of this decomposition is  then \bea
\label{deco}R(r,s) &=& \frac{\lb p\ r\rb\lb r\ q\rb\lb p\ s\rb\lb s\
q\rb}{\lb r\ s\rb^2\lb p\ q\rb^2}\left\{\frac{\half(P_z^2 Q_z^2-s_z
t_z)}{(\ell_2-k_r)^2(\ell_1+k_s)^2}-\frac{P_{L;z}\cdot
k_r}{(\ell_2-k_r)^2}-\frac{P_{L;z}\cdot k_s}{(\ell_1+k_s)^2}\right\}\nn\\
&+& \frac{\lb p\ r\rb\lb r\ q\rb}{\lb r\ s\rb \lb p\ q\rb^2}\left\{\frac{\lb s\
q\rb\lb p|P_z|r]+\lb s\ p\rb\lb q|P_z|r]}{2}\right\}\frac{\left(\frac{\eps}{1-2\eps}\right)}{(\ell_2-k_r)^2}\\
&+& \frac{\lb p\ s\rb\lb s\ q\rb}{\lb r\ s\rb \lb p\
q\rb^2}\left\{\frac{\lb r\ p\rb\lb q|Q_z|s]+\lb r\ q\rb\lb
p|Q_z|s]}{2}\right\}\frac{\left(\frac{\eps}{1-2\eps}\right)}{(\ell_1+k_s)^2}\nn
\eea where we have used the fact that \be
I^{2m}_{3:r;i}[y]=\frac{\eps}{1-2\eps}I^{2m}_{3:r;i}[1]\ee to
conveniently express the triangles' integrands.

The first coefficient above is easily recognizable as
$\frac{1}{2}b^{pq}_{rs}$ from equation (\ref{bpqrs}), but to get the
remaining two into the correct form requires an additional step.
Consider the second line of each of the four $R(r,s)$ terms.  Those
with a common value for $r$ differ only in the $s$ dependance of
their coefficients.  So when we add $R(r,m_1-1)-R(r,m_1)$, the only
change is \be \sum_{s}\frac{\lb s\ x\rb}{\lb r\ s\rb}=\frac{\lb
m_1-1\ x\rb}{\lb r\ m_1-1\rb}-\frac{\lb m_1\ x\rb}{\lb r\
m_1\rb}=\frac{\lb x\ r\rb\lb m_1-1\ m_1\rb}{\lb m_1-1\ r\rb\lb r\
m_1\rb},\ee where $x=p,q$, and we used the Schouten identity to
combine the two terms. Now the coefficient of the second line is
$\frac{1}{2}c^{pq}_{r(m_1-1)}$.  An analogous treatment of the third
line produces the coefficient $\half c^{pq}_{sm_2}$.

\subsection{Combinatorics of Cuts}

We have decomposed the integrand of each one of the MHV diagrams
into a sum (\ref{deco}) which should now be compared with  the sum occurring in the exact result
(\ref{result}). The crucial point is the decomposition of the
measure \cite{bst}: \be \frac{d^4 L_1}{L_1^2}\frac{d^4 L_1}{L_1^2}
\delta^{(4)} (L_1-L_2 +P_L) = -4 \frac{dz_1}{z_1}\,\frac{dz_2}{z_2}
\, dLIPS( \ell_2,-\ell_1, P_{L,z})\ee where dLIPS is the Lorenz
invariant phase space measure appearing in the cut rules.  For fixed
$z_1,z_2$ we have then, after performing the integration over $l_1,l_2$
sum over cuts of Feynman graphs (at shifted values of the momentum invariant).
The claim is that this sum, at $z=0$, coincides exactly with the cut of the exact result
(\ref{result}). This is not true diagram by diagram, rather there is some re-arrangement
of the cuts which we now demonstrate.

Having completed the decomposition of $R=\sum_{r,s}R(r,s)$ in the
previous section, we find it contains eight distinct terms which
are: 4 (cut) two-mass finite box functions (1 for each pair of null
legs $k_r$ and $k_s$), and 4 modified (cut) two-mass triangles (1
for each case where $k_r$ or $k_s$ is the null leg).  When we cut
the loop in the MHV diagram, this is equivalent to cutting the boxes
and triangles, as shown in figure 5, so as to keep
$\{k_{m_1},\ldots, k_{m_2}\}$ on the same side of the cut. Clearly,
which lines get cut depends completely on where $k_{m_1},k_{m_2}$
are in relation to $k_r,k_s$.  We stress that all these cuts are in
the same channel, $s$. Alternatively, we could combine the
contributions from different MHV diagrams (with different $m_1,m_2$)
which have the same null legs $k_r,k_s$ and therefore must produce
the same boxes and triangles. Different MHV diagrams will lead to
different cuts. In this manner, we may combine: the 4 boxes with
common $k_r$ and $k_s$, with cuts in the channels $s,t,P^2,Q^2$, the
2 triangles with common $k_r$, with cuts in the channels
$s=\widetilde{Q}^2$ and $P^2$, and the 2 triangles with common
$k_s$, with cuts in the channels $s=\widetilde{P}^2$ and $Q^2$, for
all values of $r,s$.

\begin{figure}[p]
\begin{center}
\epsfbox{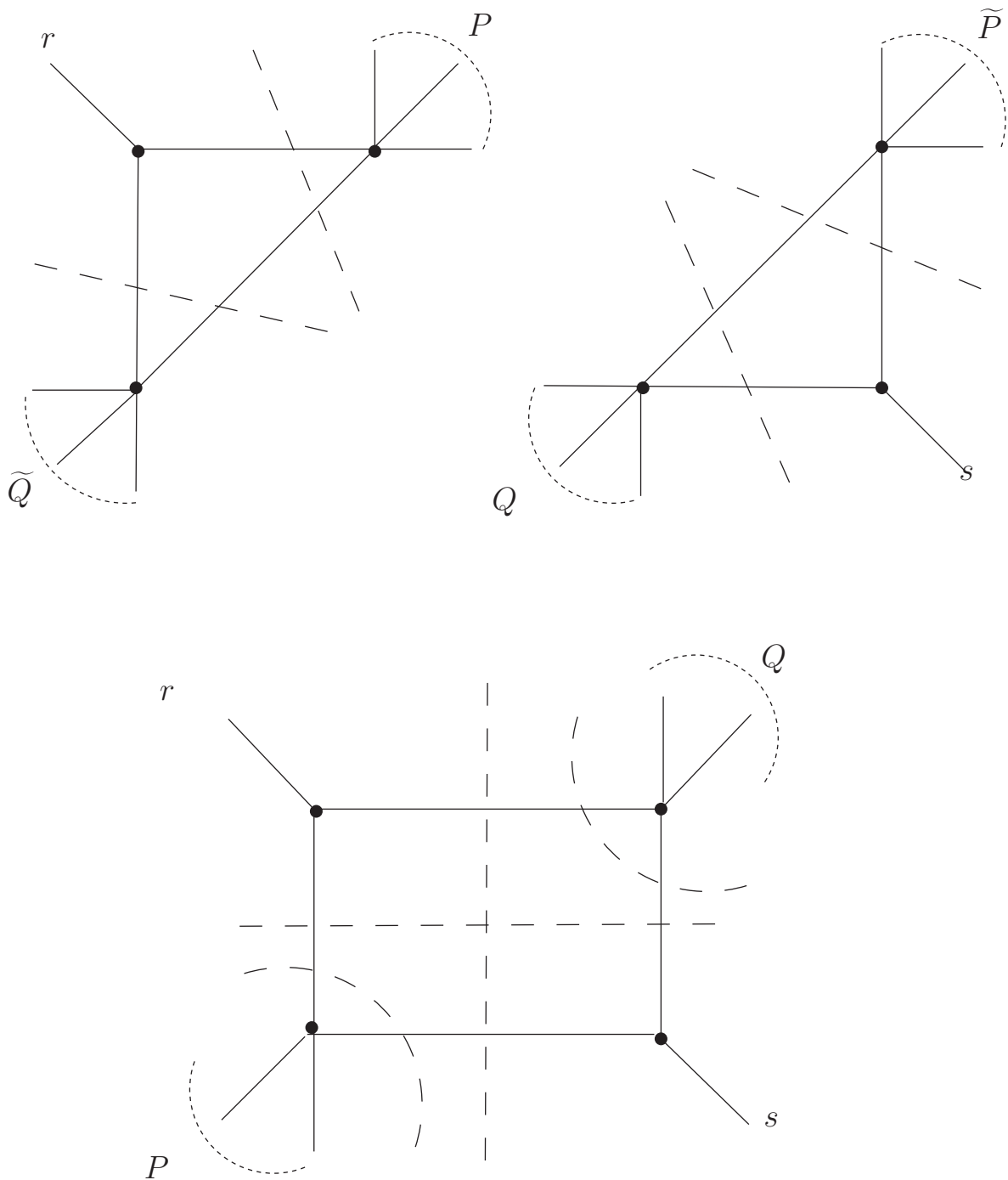}
         \end{center}
         \caption{
         One MHV diagram produces 4 cut boxes and
         triangles, one for each dashed line. Where exactly the cut lies
         depends on $r,s$.  Therefore, a given box (triangle) with
         $r,s$ fixed requires 4 (2) MHV diagrams to produce all of its cuts.}
\end{figure}

In the exceptional cases where one of the triangles massive legs
becomes massless, then this diagram has the single non-trivial cut
which isolates the remaining massive leg. We will show below that
each of these terms are reconstructed from their single cut.

One might worry that not all the cuts exist in all channels for
non-degenerate cases. A priori, we must sum over all MHV diagrams
with $q+1\leq m_1\leq p$ and $p\leq m_2\leq q-1$, but when $m_2=p$
or $m_1-1=q$ the corresponding boxes and triangles may not be
defined. Fortunately, the coefficients \be
b^{pq}_{ps}=b^{pq}_{rq}=c^{pq}_{ps}=c^{pq}_{qr}=0\ee all vanish. So,
we may restrict the sums over $m_2\equiv r,m_1\equiv s+1$ to the
ranges given in Section 2.2, plus the degenerate triangle terms.

So, in summary, we have found that the decomposition of the sum of
MHV diagram is simply related to the result (\ref{result}). For any
channel $X=s,t,P^2,Q^2$, of any function $F=B,T,\A_{IR}$ in
(\ref{result}), we find a term in our sum of the form $\Delta_X
F(X_z)$, where $\Delta_X$ denotes the cut in the $X$-channel, and
$X_z$ is $X$ shifted by $z$-dependent terms.

\subsection{Calculating the Cuts}

In the last section, we noted that the loop integrations factor into
two parts: dispersion integrals over the $z_i$ and an integral over
dLIPS$(\ell_2,\, -\ell_1,\, P_{L;z})$ which computes the cuts in the
diagrams.  The cut box integrals were computed in \cite{bst}, so the
only new ingredients are the cut triangles.  We will now
evaluate these integrals for when $k_r$ is the null leg, the other
case follows by switching $r\leftrightarrow s$ and
$\ell_2\leftrightarrow -\ell_1$.  Also, we focus on the $s$-channel
cuts; other channels are treated analogously.  The integrals we
wish to solve are in dimension $D=4-2\eps$ and of the form \be
\mathcal{I}(s_z)=\int
d^{D}\mbox{LIPS}(\ell_2,-\ell_1,P_{L;z})\frac{N(P_{L;z})}{(\ell_2-k_r)^2},\label{cut}\ee
where the numerator $N(P_{L;z})$ only depends on $P_{L;z}$ and
external momenta\footnote{Since [r r]=0, we can always write $\lb
x|P_z|r]=\lb x|(P_z+k_r)|r]=\lb x|P_{L;z}|r]$, where $x=p,q$.}.  By
boosting to the rest frame of $\ell_1-\ell_2$, then rotating $k_r$
into the $x_D$ direction, we have \be
\ell_1=\half\,|P_{L;z}|\,(1,\textbf{v})\ ;
\qquad\ell_2=\half\,|P_{L;z}|\,(-1,\textbf{v})\ ; \qquad
k_r=(k_r,0,\ldots,0,k_r),\label{frame}\ee where the unit vector
$\textbf{v}$ is such that $\textbf{v}\cdot
\widehat{x}_{D}=\cos(\theta_1)$.  This allows us to re-write our
phase-space measure as in \cite{bst} \be
d^{D}\mbox{LIPS}(\ell_2,-\ell_1,P_{L;z})=
\frac{\pi^{\half-\eps}}{4\Gamma(\half-\eps)}
\left|\frac{P_{L;z}^2}{4}\right|^{-\eps}
d\theta_1d\theta_2(\sin\theta_1)^{1-2\eps}(\sin\theta_2)^{-2\eps}\ee
and the integrand's denominator becomes \be
(\ell_2-k_r)^2=-2\ell_2\cdot
k_r=k_r\,|P_{L;z}|\,(1-\cos\theta_1).\ee Performing the integral
(\ref{cut}) is now a simple task, with the result \bea \label{cuts}
\mathcal{I}(s_z) &=&
\frac{4^{\eps}\pi^{\frac{3}{2}-\eps}}{2\Gamma(\half-\eps)}
\left|\frac{s_z}{4}\right|^{-\eps}\frac{N(P_{L;z})}{k_r|P_{L;z}|}\frac{\Gamma(-\eps)}{\Gamma(1-\eps)}\nn\\
&\rightarrow&-\frac{1}{\eps}\frac{\pi}{2}\,\frac{N(P_{L;z})}{k_r|P_{L;z}|}\,s_z^{-\eps}.\eea

Now, for any channel of any function $F(X)$ appearing in the result
(\ref{result}), we are left with an integral of the form $\int
\frac{dz_1 \ dz_2}{z_1 \ z_2} \Delta_X F(X_z)$, where the cuts of the
triangle graphs are exhibited in (\ref{cuts}). Furthermore, we have
shown that $\Delta_X F(X_{z=0})$ is precisely the cut of the exact
result (\ref{result}). Appealing to cut constructibility, we can
anticipate that our dispersion integration will reproduce the
correct answer as long as the functions $\Delta_X F(X_z)$ are cut
free on the integration contour of the $z$-integration.
 As the cuts (\ref{cuts}) do include non-analytic functions of $X_z$, the correct
 contour\footnote
 {We note that the integration contour then is channel-dependent, as in \cite{bst}.} is
$X_z \geq 0$.  Choosing such contour, it is a simple matter to perform the dispersion
integration directly to verify that we get the correct answer, and we turn to that
integration now.

\subsection{Dispersion Integrals}

We now show that the  final integrations over $z_1,z_2$ reproduce
the result in (\ref{result}).  Recall that in section 3.2, we
demonstrated that the sum over MHV diagrams is equivalent to the sum
of cuts in all possible channels of the box and triangle diagrams.
Thus, it remains to show that a given box (triangle) is
reconstructed from the sum of its 4 (2) integrated cuts.

We change our integration variables to $z\equiv z_1-z_2,\ z'\equiv
z_1+z_2$ and note that for any function $f(z)$ independent of $z'$
\be \int \frac{dz_1}{z_1}\frac{dz_2}{z_2}f(z_1-z_2)=2(2\pi i)\int
\frac{dz}{z}f(z).\ee  Next, we use the fact that $s_z=s-2z\eta\cdot
P_{L}$ to write \be \frac{dz}{z}=-\frac{ds_z}{s-s_z},\ee with a
corresponding change of variables in the other channels.  Now, we
must show that \bea\label{sumbcuts} B(k_r,Q,k_s,P) &=&
\int\frac{ds_z}{s-s_z}\Delta_s B(s_z)+\int\frac{dt_z}{t-t_z}\Delta_t B(t_z)\nn\\
&-&\int\frac{dP^2_z}{P^2-P^2_z}\Delta_{P^2}
B(P^2_z)-\int\frac{dQ^2_z}{Q^2-Q^2_z}\Delta_{Q^2} B(Q^2_z)\eea and
\be\label{sumtcuts}
T(k,P,Q)=\int\frac{dP^2_z}{P^2-P^2_z}\Delta_{P^2}
T(P^2_z)-\int\frac{dQ^2_z}{Q^2-Q^2_z}\Delta_{Q^2} T(Q^2_z).\ee

Again, we will consider the $s$-channel only, the other channels
follow immediately.  As discussed above, we must choose integration contour for which
the integrands are analytic as functions of the kinematical variables. Therefore  we
 restrict the
integration to $s_z>0$, where the expression (\ref{cuts}) has no cuts.

First, we will reconstruct the divergence free box functions.  They
posses three types of terms, given in the first line of
(\ref{deco}). The first of these was calculated in \cite{bst}, we
quote their result: \be \int\frac{ds_z}{s-s_z}\frac{\half(Q^2_z
P^2_z-s_z
t_z)}{(\ell_2-k_r)^2(\ell_1-k_s)^2}=-\frac{1}{\eps^2}(-s)^{-\eps}-\mbox{Li}_2(1-a\,s)\ee
The next term has the cut $\mathcal{I}(s_z)$ found earlier, with
numerator $N(P_{L;z})=-P_{L;z}\cdot k_r$.  Up to a sign, this
numerator is precisely the denominator in our working reference
frame (\ref{frame}).  The dispersion integral is then \be
\label{dispint} -\frac{1}{2\eps}\int_0^\infty\frac{ds_z}{s-s_z}=
\half\frac{\pi\csc(\pi\eps)}{\eps}(-s)^{-\eps}\longrightarrow\frac{1}{2\eps^2}(-s)^{-\eps}\ee
The next term in the divergence free box gives an identical
contribution.  Summing the three contributions, we find \be
\int\frac{ds_z}{s-s_z}\Delta_s B(s_z)=-\mbox{Li}_2(1-a\,s),\ee
exactly what is required to reproduce (\ref{box}).  Treating the other
channels similarly proves the equality of (\ref{sumbcuts}) and
(\ref{box}).

Moving on to the triangles, we will consider those where $k_r$ is
the null leg.  These also have cuts of the form $\mathcal{I}(s_z)$,
in the reference frame (\ref{frame}) the numerator is \be
N(P_{L;z})=\lb x|P_z|r]=\lb x|P_{L;z}|r]=\lb
x|\gamma^0|r]\,|P_{L;z}|\ee times $(\frac{\eps}{1-2\eps})$ and
$x=p,q$. The dispersion integral is nearly identical to
(\ref{dispint}): \be
\frac{1}{2\eps}\int_0^{\infty}\frac{ds_z}{s-s_z}\frac{\eps}{1-2\eps}\frac{\lb
x|\gamma^0|r]}{k_r}s_z^{-\eps}=\frac{1}{\eps(1-2\eps)}\frac{\lb
x|\gamma^0|r]}{2k_r}(-s)^{-\eps}.\ee  Multiplying the top and bottom
by $|P|\equiv P^0$, then re-expressing this result in a covariant
fashion gives the coefficient \be \frac{\lb
x|\gamma^0|r]}{2k_r}=\frac{\lb x|P|r]}{2k_r\cdot P}=\frac{\lb
x|P|r]}{\widetilde{Q}^2-P^2}\ee (recall that
$s\equiv\widetilde{Q}^2$).  An analogous result holds in the $P^2$
channel.  Taking the difference of the two, and expanding
$(-\widetilde{Q}^2)^{-\eps},(-P^2)^{-\eps}$ in $\eps$ yeilds the
desired result: \be \frac{1}{\eps(1-2\eps)}
\frac{(-\widetilde{Q}^2)^{-\eps}-(-P^2)^{-\eps}}{\widetilde{Q}^2-P^2}
=\frac{\log(\widetilde{Q}^2)-\log(P^2)}{\widetilde{Q}^2-P^2}.\ee

In the case of the one-mass triangles, the result is even simpler.
 Consider the case $(r,s)=(p+1,p-1)$, then $P^2=p^2=0$ and the dispersion integral
gives \be
\frac{1}{\eps(1-2\eps)}\frac{(-\widetilde{Q}^2)^{-\eps}}{\widetilde{Q}^2}\ee
as desired. We conclude therefore that  all triangle terms
 are reconstructed once we perform the final dispersion integration.

Thus, we have shown by explicit calculation that the MHV diagrams formalism is valid for
the calculation of one-loop contribution of the $\mathcal{N} =1$ chiral
multiplet to the MHV amplitude. Together with the result of \cite{bst} this establishes
the validity
of the MHV-diagram technique for that amplitude in any massless supersymmetric theory. It would be interesting to check
the formalism
further by applying it to non-supersymmetric, but cut-constructible amplitudes.

\section*{Acknowledgements}
We thank Zvi Bern for a useful comment. M.R. thanks the physics
departments at Cornell and Syracuse, and the KITP at Santa Barbara
for hospitality   during the completion of this work. We are
supported by National Science and Engineering Research Council of
 Canada. This research was supported in part by the National Science Foundation under
Grant No. PHY99-0794

\end{document}